\documentstyle[11pt,paspconf,epsf]{article}

\begin{document}

\title{The stream-disk interaction}
\author{Philip J. Armitage}
\affil{Canadian Institute for Theoretical Astrophysics, McLennan Labs,
       60 St George St, Toronto M5S 3H8, Canada}

\begin{abstract}
I review theoretical aspects of the interaction between the accretion 
stream and the disk in interacting binary systems, concentrating on 
recent hydrodynamic calculations. At low accretion rates, cooling is
expected to be efficient, and the interaction leads to a nearly ballistic
stream overflowing the disk rim towards smaller radii. If cooling is 
ineffectual, the shocked gas produces a bulge on the disk rim, and there
is no coherent stream inward of the disk edge. Results are presented for
the mass fraction and velocity structure of the overflowing component, 
and the implications for X-ray observations of `dips', and doppler 
tomography of cataclysmic variables are briefly discussed.
\end{abstract}


\section{Introduction}
Mass transfer via Roche lobe overflow onto an accretion disk occurs
in a diverse range of non-magnetic and weakly magnetic cataclysmic
variables, low-mass X-ray binaries, and supersoft X-ray sources. In 
all of these systems an accretion stream from the inner Lagrange point 
strikes the outer regions of the disk in a highly supersonic impact,
producing most obviously a hot spot at the disk edge. However it has long been
realised that the impact can also have interesting secondary consequences
(for a review, see e.g. Livio 1993). 
For example the stream material may be able to overflow the disk rim
(Lubow \& Shu 1976; Frank, King \& Lasota 1987) or the interaction might 
throw stream gas high above the disk plane. Either possibility could well 
lead to marked departures from axisymmetry in both the absorption towards the
central object (producing `dips' in the X-ray and UV light curves; 
Mason, 1989; Hellier, Garlick \& Mason 1993), and in the
structure of a disk wind. 

In this contribution I discuss first the dynamics of the stream flow up 
to the point of impact with the disk, following the approach of Lubow \& 
Shu (1975, 1976), and Lubow (1989). This analysis
demonstrates that the stream is potentially more extended vertically 
than the disk at the point of impact and can thus overflow the disk rim -- and 
additionally serves as initial conditions for three-dimensional 
hydrodynamical calculations, using the ZEUS code, that are presented 
later. The use of such simulations is necessary to explore fully the
complex structure of the impact region, and the subsequent interaction 
of stream and disk material. Here I analyze the simulation results 
with a view towards the observable quantities -- the mass fraction 
of stream gas that overflows, and the velocities attained by the
overflowing component.

\section{Stream flow dynamics}

\begin{figure}
\plotone{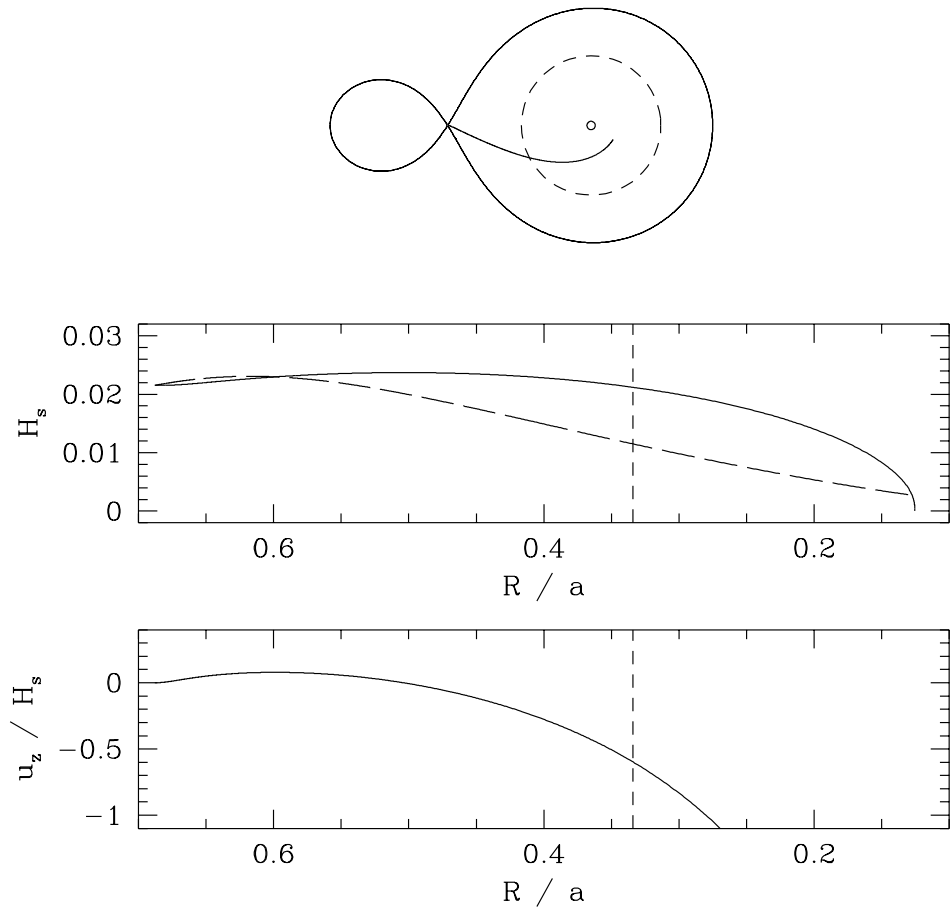}
\caption{Results of a vertical integration of the stream dynamics,
following Lubow (1989), showing (top) the path taken by the gas stream in the
Roche potential. The upper panel shows the scale height of the stream (solid 
line) 
as a function of the fractional radius $R/a$, where $a$ is the 
binary separation. The thickness expected if the 
stream were in local hydrostatic equilibrium is shown as the dashed line.
The lower panel depicts the vertical velocity of the stream at one stream 
scale height, in units of the stream sound speed. The dashed circle and 
vertical lines denote the observed disk radius in Z Cha. At the disk edge, 
the stream is (i) thicker than its hydrostatic value, and (ii) converging
on the disk at moderately supersonic velocities at a few scale heights
above the midplane.}
\end{figure}

A detailed analysis of the formation and dynamics of an accretion 
stream arising from Roche lobe overflow is given by Lubow \& Shu 
(1975, 1976). The stream scale height, $H_{\rm s}$, on leaving the 
vicinity of the inner Lagrange point is given by,
\begin{equation}
 H_{\rm s} = {c_s \over \omega_z}, \ \ \ \ \ \ \ \omega_z^2 = 
 - { {\partial^2 \phi} \over {\partial z^2} } 
 \vert_{x_{L_1},y_{L_1}}
\end{equation}
where $c_s$ is the sound speed at the surface of the mass losing 
star and $\phi$ the gravitational plus binary centrifugal potential. 
To order of magnitude the stream thickness is $\sim c_s / \Omega_{\rm B}$,
where $\Omega_{\rm B}$ is the binary angular velocity (Livio 1994). Note 
that this is much larger than the pressure scale height at the surface 
of the secondary. Once the stream has left the vicinity of $L_1$,
it flows inward on an essentially ballistic trajectory until it 
meets either the disk edge or the magnetosphere of the accreting 
star. However, because the flow inward occurs on a dynamic timescale, 
the stream is unable to adjust its vertical structure fast enough 
to maintain hydrostatic equilibrium. This inertia effect,
which is analyzed in Lubow \& Shu (1976) and applied to the dwarf
nova Z Cha by Lubow (1989), makes the 
stream thicker than it would otherwise be. 

Figure 1 shows the results of applying this analysis to a system
like Z Cha (mass ratio $1/7$). The stream
when it reaches the edge of the disk is much thicker than it would 
be if it were in local hydrostatic equilibrium, and thus although the
{\em central} density of the stream is much less than that of the 
disk (as for the disk $v_R \ll v_\phi$), at a few scale heights
above the plane the stream density may exceed that of the disk. 
The upper part of the stream may thus be able to overflow the 
rim of the disk, before collapsing onto the disk surface, 
at a phase of $\sim 0.5 - 0.6$ -- almost independent of mass
ratio.

This analysis clearly establishes the potential for stream 
overflow to occur, but cannot address either the interaction
of the stream as it flows over the disk surface, or the 
possible hydrodynamic effects at the impact point that might
disrupt the stream. 

Even within the context of this simplified picture, 
stream overflow is {\em not} expected for all
systems, at all times. Since the relevant comparison is the 
scale height of the stream as compared to the disk at the disk
edge, overflow will be favored if the disk is thin -- i.e. if the
temperature at the edge is low compared to that of the secondary.
This will favor large disks, with low accretion rates. Conversely, 
shorter period systems, with small disks, or systems in outburst 
where $\dot{M}$ is large, are unlikely to allow simple ballistic 
stream overflow.

\section{Hydrodynamics of the stream-disk interaction}

Simulations of the stream impact with a disk in a binary potential 
were presented by Armitage \& Livio (1996). In those calculations, 
which employed smooth particle hydrodynamics with an isothermal 
equation of state, the dominant cause of non-axisymmetric 
structure at high elevation was found to be stream overflow. 
Additional non-axisymmetric structure arose from the eccentricity 
of the disk, though this would only be visible in {\em absorption} 
for viewing angles rather close to the disk plane (for a discussion 
of the emission from such distorted disks, see e.g. Murray, this
volume). However the resolution of the impact region in
the SPH calculations was rather poor, and insufficient for a
detailed comparison with either observations or the Lubow (1989)
analysis.

To investigate the impact region and subsequent stream-disk interaction 
in more detail, we have used the ZEUS-3D hydrodynamics code developed at 
the Laboratory for Computational Astrophysics. ZEUS-3D is a finite difference 
code that uses an artificial viscosity to capture shocks, for details 
see the papers by Stone \& Norman (1992a,b) that describe the earlier,
two-dimensional, version of the code.

\subsection{Initial conditions}

The simulations cover a computational box surrounding the impact point,
with the disk and stream inflow prescribed as boundary conditions. For 
the disk, we assume a hydrostatic vertical structure with a Mach number 
at the disk edge of 30. The disk is isothermal both vertically and 
radially, though relaxing the latter requirement does not greatly 
alter the results. Beyond the disk `edge' the density is assumed to 
fall off as a gaussian, with radial scale length equal to the vertical
scale height.

For the stream, we use initial conditions akin to those obtained from 
the vertical integration presented in the preceeding Section. The 
stream density is assumed to fall off as a gaussian, with a vertical 
velocity $v_z \propto -z$. The Mach number of the stream at the disk 
edge is 30, and the stream makes an angle of $15^\circ$ with the 
radial direction. The ratio of disk to stream central density is 100.
We vary the assumed ratio of stream to disk scale heights, 
$H_{\rm s} / H_{\rm d}$, and the equation
of state, here either isothermal or adiabatic. Simulations including
optically thin radiative cooling are generally similar to the isothermal 
runs for all reasonable accretion rates. Further details of the 
calculations are discussed elsewhere (Armitage \&  Livio 1997).
Note that we do not include any of the effects (such as irradiation 
of the disk by a central X-ray source) that distinguish CVs from 
X-ray binaries or supersoft sources, see Blondin (1997) for a
description of simulations specifically modelling X-ray binaries.

\subsection{Cooling and the effective equation of state}

For accretion rates normally encountered in cataclysmic variables,
the density of stream and disk gas is sufficiently high that 
optically thin cooling is rapid (i.e. fast compared to the dynamical 
time) in the disk midplane. In this regime of rapid optically thin
cooling, the general hydrodynamics is similar to an isothermal 
equation of state (see Blondin, Richards, \& Malinowski, 
1995, for a discussion of this applied to simulations of
accretion in Algol). At higher accretion rates, however, the 
emission from the shocked gas will cease to be optically thin, 
and we expect that the hydrodynamics will be `more like' that seen 
in an adiabatic simulation, though there are many complexities 
that such an assumption ignores. 

To estimate the critical accretion rate where the hot spot emission
first becomes optically thick, we consider the opacity both of the 
cold inflowing stream gas, probably arising from H$^{-}$, and the 
hot shocked layer where the opacity is assumed to be from electron
scattering. With large uncertainties, either opacity suggests that
$M_{\rm crit} \sim 10^{-9} \ M_\odot {\rm yr}^{-1}$ (Armitage \& Livio 
1997). Although very crude, this suggests that the hotspot region in 
low accretion rate CVs may well be able to cool efficiently, whereas
nova-like and super-soft X-ray sources with much higher accretion rates
are almost certainly unable to do so. The former are possibly better 
described by the isothermal calculations, the latter by adiabatic 
simulations. 

\subsection{Results}

\begin{figure}
\plotone{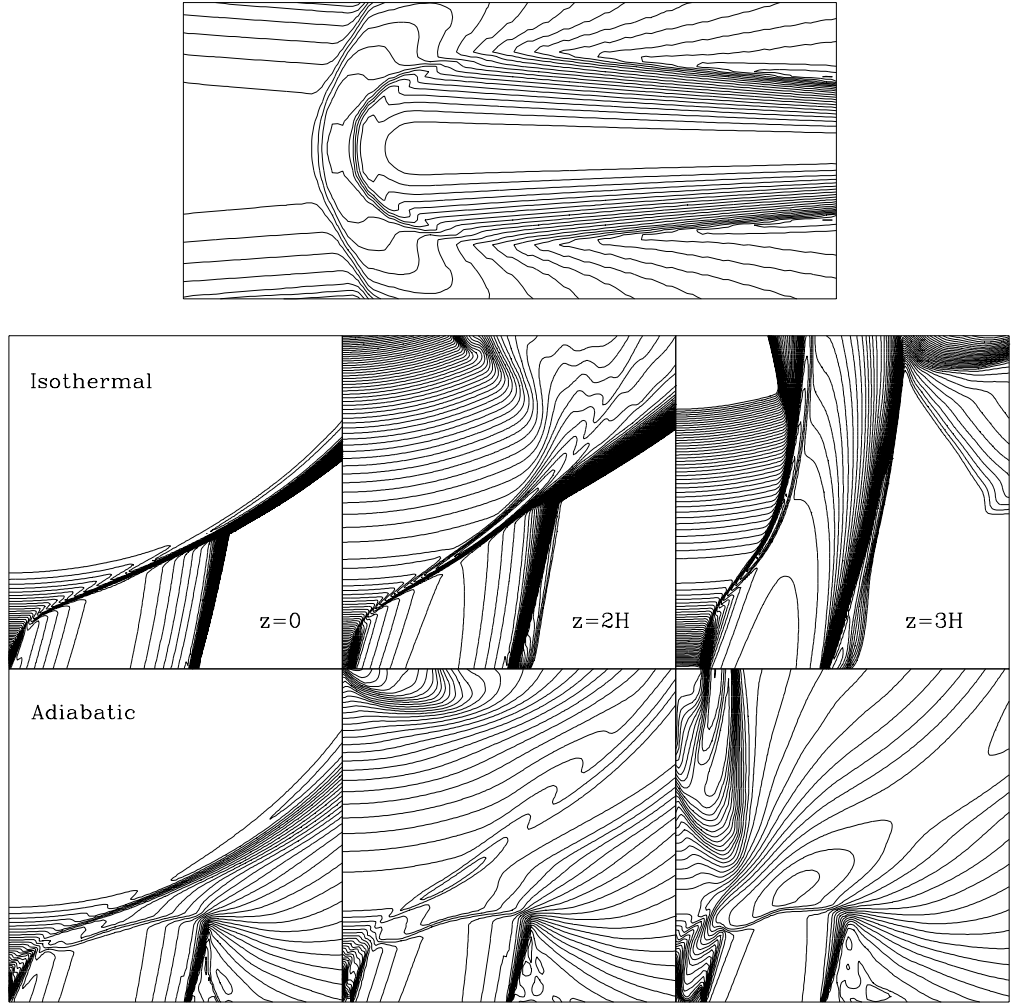}
\caption{Flow in a hydrodynamic calculation of the stream-disk interaction
(Armitage \& Livio 1997).
The upper panel is a slice along the initial stream flow direction for 
an adiabatic equation of state, the lower panels are slices 
parallel to the disk 
midplane at $z=0,2,3$ {\em disk} scale heights $H$, for both isothermal
and adiabatic calculations. The stream enters from the bottom left, the
disk is flowing left to right across the frame which covers $\sim$ 50 \%
of the disk radius.
Density contours are plotted at 
$\Delta \log \rho = 0.2$.}
\end{figure}

Figure 2 shows results from an isothermal calculation with $H_{\rm s} / 
H_{\rm d} = 2$, and an adiabatic run with $H_{\rm s} / H_{\rm d} = 2.5$.
For the isothermal run, the stream is stopped rapidly in the disk 
midplane by the denser disk gas. Moving above the midplane, at around
2 disk scale heights the stream and disk densities become comparable,
and we resolve the twin shocks in the disk and stream gas seen 
in earlier two dimensional calculations (Rozcyczka \& Schwarzenberg-Czerny 
1987). However the 3D calculations do not reproduce a hot spot region 
highly extended along the disk rim, as was seen in 2D. Moving further upwards, 
the stream rapidly dominates. At 3 $H_{\rm d}$ the stream overflows almost
freely, with some modest deflection at the edge facing the incoming disk 
material.

The corresponding slices from the adiabatic run are very different. In the
midplane, there is pronounced `splashing' of hot material downstream of
the impact point, while at 3 $H_{\rm d}$ there is no coherent stream 
overflowing the disk, only a broad fan of material moving across the disk 
surface.
 
A slice along the stream flow direction shows a prominent bow shock in 
the adiabatic simulation, in this case material is expanding in a
wide fan downstream and above the impact point. Viewed as isodensity 
surfaces, the effect is to produce a bulge in the disk rim downstream 
of the impact, though the material in the bulge is definitely {\em not} 
close to hydrostatic equilibrium -- it is expanding and cooling rapidly. 
Conversely the isothermal simulation looks very similar to the stream 
overflow proposed by Lubow \& Shu (1976) or Frank, King \& Lasota 
(1987), though even in this case there is much more swept up disk gas 
or entrained stream material than freely overflowing material, with 
this additional gas showing intermediate inflow velocities.

\subsection{Overflowing mass fraction}

\begin{figure}[t]
\plotone{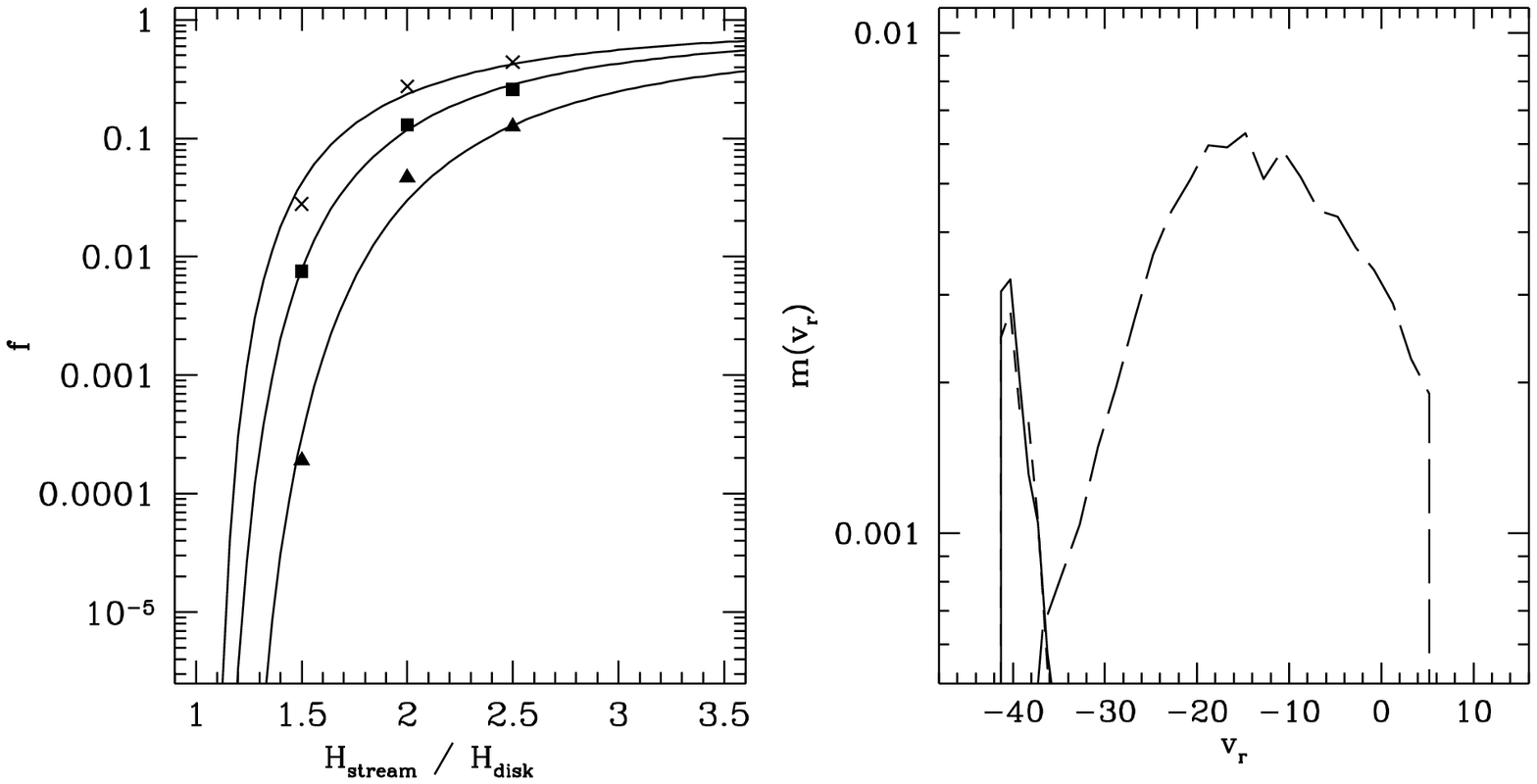}
\caption{Left panel -- the overflowing mass fraction as a function of the
ratio of stream to disk scale height, for efficient radiative cooling. The
symbols show the fraction overflowing with radial velocity greater than
$5 c_s$ (crosses), $10 c_s$ (squares) and $20 c_s$ (triangles). The
curves are a fit described in the text. Right panel -- the velocity 
of material along a line of sight to the primary $32^\circ$ downstream
of the impact point, and $12^\circ$ above the disk midplane. The solid
line is the result for the efficient cooling simulation -- almost 
identical to the free-stream model shown as the dashed line. The long
dashed line is the adiabatic result, more mass overflows in this case
and it has a much broader range of velocities.}
\end{figure}

For the simulations with efficient radiative cooling, Figure 3 shows the
fraction of stream gas that is overflowing with $|v_R| > v_{\rm cut}$ as
a function of $H_{\rm s} / H_{\rm d}$. Results
from the simulations have been fit with 
curves representing the most naive assumption -- that {\em all} of the 
stream with $z$ greater than some $z_{\rm crit}$ overflows, while
everything below $z_{\rm crit}$ is stopped on impact. For $z_{\rm crit}$
we take, 
\begin{equation}
 z_{\rm crit} = \beta { {H_{\rm d} H_{\rm s} } \over 
 \left( H_{\rm s}^2 - H_{\rm d}^2 \right)^{1/2} }
 \left( \ln { \rho_{\rm d} \over \rho_{\rm s} } \right)^{1/2},
\end{equation}
where $\rho_{\rm d}$, $\rho_{\rm s}$ are the midplane disk and stream densities.
$z_{\rm crit}$ is the height where the stream and disk densities are equal, 
multiplied by a factor $\beta$ included as a free parameter.

From the Figure, it is evident that this fitting formula, with $\beta$ 
close to unity, provides a fair estimate of the amount of highly 
supersonic stream overflow i.e. roughly as much material overflows the 
rim as one would naively expect. This implies that $H_{\rm s} / H_{\rm d}$ 
must be around 2 or greater for the overflowing fraction to be reasonably
large. Note also that there is much more material overflowing at lesser,
but still highly supersonic, velocities than there is ballistic stream
gas.

Despite the striking qualitative distinction, a similar (or rather larger) 
mass fraction overflows the disk in the adiabatic simulation. In this 
case though the material extends to much higher above the disk midplane,
and of course has a much broader range of velocities. 

\subsection{Velocity profiles -- a first look}

The velocity structure of the stream gas after reaching the disk 
edge is potentially observable via doppler mapping techniques. 
For example the structure of the disk rim near the hot spot has been 
considered in OY Car by Billington et al. (1996), and for SW Sex by 
Dhillon, Marsh \& Jones (1997). We have not, yet, attempted to 
generate synthetic doppler maps from the simulations, but Figure 3
provides some idea of the velocity profile of gas along a line of 
sight to the primary. Plotted are the results for isothermal and
adiabatic simulations, and a comparison `free-stream' run which 
included the internal stream pressure effects but omitted entirely 
the disk. 

At high angles above the disk midplane, the velocity profiles for 
the isothermal and free-stream calculations are essentially identical,
implying that for studies of absorption at high angles a ballistic 
stream approximation is probably adequate (if the cooling is 
indeed efficient). However closer to the 
disk plane there is a large contribution from entrained material 
with a lesser radial velocity -- this might well appear in emission
as the stream reaches the point where it reimpacts the disk (see, e.g. 
Hellier 1996). As expected from 
the appearance of the flow, inefficient cooling leads to a much 
broader range of line-of-sight radial velocities, which are generally
of smaller magnitude than for the isothermal simulation. Of course the
observed profile will strongly depend on the degree of cooling that 
{\em does} take place as the shocked gas expands -- we have here 
considered only the limit of no radiative cooling at all. Nonetheless the 
prediction is that inefficient cooling is likely to lead to absorption
with a broad range of radial velocity, and be more significant at 
larger angles from the disk plane.

\acknowledgments
My thanks to Mario Livio for his advice on all aspects of the work
presented here, to Norm Murray for useful discussions, and to the
conference organizers for a highly enjoyable meeting.

\end{document}